\documentclass[a4paper,11pt]{article}
\usepackage{latexsym}
\usepackage{epsfig, subfigure, subfloat, graphicx, float }
\usepackage{amssymb}
\usepackage{amsmath}
\usepackage{epstopdf}
\usepackage{authblk}
\usepackage{amsmath}
\title{Hairy Waves:\\Gravitational Waves in the presence of Screening Mechanism}

\author[1]{M. Honardoost\thanks{m\_honardoost@sbu.ac.ir}}
\author[1,2]{N. Khosravi \thanks{n-khosravi@sbu.ac.ir}}
\author[1]{N. Rahmanpour\thanks{n\_rahmanpour@sbu.ac.ir}}
\affil[1]{\small Department of Physics, Shahid Beheshti University, G.C., Evin, Tehran
19839, Iran}
\affil[2]{\small School of Physics, Institute for Research in Fundamental Sciences (IPM), P.O. Box 19395-5531, Tehran, Iran}

\begin{document}

\maketitle
\begin{abstract}
We consider a binary system in the context of screened modified gravity models and investigate its emitted gravitational wave which has been shown that is a hairy wave. We derive power of the gravitational wave and the binding energy of the binary system and then trace back the effect of the additional scalar field, present at screening models, in the frequency of GW emitted from the binary. The binary is considered as two astrophysical black holes in a spherically symmetric galactic ambient/halo with constant density. Effectively in the ISCO region the screening force is in charge  due to the sharp transition in the density of matter field. It is shown that, in the presence of screening, when one companion of the binary system is between $R_{thinshell}$ and $R_{ISCO}$, the frequency of the gravitational emission have decreased, compared to  GR without screening.

\end{abstract}

\section{Introduction}
Recently, a 100 years old prediction of Einstein general relativity (GR) has been approved by direct detection of gravitational waves (GW) by LIGO \cite{ligo}. This event is a milestone in the research field of gravitation. The sources of the detected GWs are binary black holes (BH-BH) and one binary neutron star (NS-NS). The NS-NS event is very informative since in addition to GW, its electromagnetic counterpart has been detected. The comparison between GW and electromagnetic wave constrains the existence of any modification of Einstein-Hilbert gravity. Basically this result is a consequence of almost same speed of propagation for GW and electromagnetic wave \cite{paolo}. However it should be emphasized that this result is true under assuming certain conditions e.g. there is no screening mechanism. By construction, existence of screening mechanism screens the effects of any modification to gravity. Therefore, we may expect that the GW constraints, based on the speed of propagation,  cannot be applied to modified gravity models with screening mechanism.

The idea of screening mechanism is that gravitational force depends on the environment. Indeed, in this mechanism, gravity force depends on the density of the matter fields. By using this mechanism modified gravity theories are restored to GR in small scales. Chameleon\cite{cham}, symmetron\cite{symmetron} and Vainshtein\cite{vainsh} are most famous examples of screening mechanism in scalar-tensor theories which the additional scalar degree of freedom is screened in high density regions. Chameleon mechanism can be realized as if the scalar degree of freedom is coupled to the matter field in such a way that this scalar be short-ranged. In symmetron mechanism, the coupling between matter and scalar turns off in dense environment while scalar couples to matter in low density regions. In Vainshtein mechanism, on the other hand, the scaler degree of freedom is suppressed dynamically by the kinetic term of the action. The various aspects of these models have been widely studied in cosmology \cite{smcos}. In this manuscript, we work with screened modified gravity (SMG) and it just refers to a family of screened scalar-tensor theories wherein the scalar potential in the action is responsible for the screening behaviour. In SMG models matter field is non-minimally coupled to the scalar field in Einstein frame.

It should be noted that the effect of this non-minimal coupling on astrophysical objects like neutron star was investigated by Damour and Esposito-Farese (DEF) \cite{damour,damour2}. It is shown that on the certain conditions, there exists a critical baryonic mass above which the star posses a non-zero scalar profile \cite{scalarization}. This process, which is named as scalarization, leads to a deviation from GR inside and in the vicinity of the star, while safely passes the solar system experiment \cite{damour2,scalarization}. In this case, gravity is weaken inside the neutron star and effective gravitational constant decreases in the spontaneous-scalarization phase \cite{sccal1}.

Similar process to the DEF can also occur for some types of black holes. For example, it is shown that in the case of matter surrounding Kerr black holes in scalar tensor theories, the scalar field would excite and a non-trivial scalar profile can emerge \cite{vitor}. In other words, hairy black holes emerged in scalar tensor theories with non-minimal coupling. Therefore, one can deduce that the emitted GW from hairy binary black hole should be different from its GR counterpart. Accordingly, A. C. Davis et al. in \cite{ac1,ac2} have investigated the emitted GW from hairy black hole in SMG, as these models are in fact a special type of scalar-tensor theories with non-minimal coupling. They presented an analytical and numerical description of these black holes and showed that how different the behaviour of GW solutions are in comparison with GR. Note that the authors have compared the effect of scalar interaction in GW by a classical estimation.

In this paper, we follow \cite{ac1} with some differences in approach. We study a binary system in the context of SMG models without investigation of the scalar profile. For this purpose, we evaluate the binding energy of the binary system in SMG models and trace back the effect of the fifth force (the scalar field) in the frequency of GW emitted from the binary. In addition, we consider just the cases in SMG models wherein no scalar mode exists. It is evident that this condition put new constraints on the mass of scalar field in comparison with local test of gravity. It is worth noting that the environment of a binary system (either BH-BH or NS-NS) can be an interesting region to study screening effects due to the existence of innermost stable circular orbit (ISCO). In fact, since ISCO region is almost empty, there is a sharp change in the density of matter field in this region of space. In other words, the screening mechanism is very effective in ISCO region and its effects cannot be negligible if there are any.

In the following sections, we briefly review SMG models and a BH-BH system in this context. Then we study the GW emitted of the binary and derive power of it. By comparison of the power with the energy loss of the spiral, frequency of the gravitational emission is derived. Therefore, the effects of screening mechanism in the form of this frequency can be compared with its counterpart in GR.

\section{Screened modified gravity}

Let's consider
\begin{equation}\label{1}
  S=\int d^{4}x \sqrt{-g} [\frac{M^{2}_{pl}}{16\pi}\mathcal{R}-\frac{1}{2}(\nabla \phi)^{2}-V(\phi )] + S_{m}[\psi , A^{2}(\phi)g_{\mu \nu}],
\end{equation}
as a prototype SMG action. In this action, metric $g_{\mu \nu}$ and scalar $\phi$ are dynamical variables and $\psi$ is matter field. $g$ is the determinant of the metric, $M_{pl}$ is the reduced Planck mass and $\mathcal{R}$ refers to the Ricci scalar in Einstein frame. The function $V(\phi)$ is the bare potential of scalar field and without loss of generality we assume a component of matter field $\psi$ couples through $A^{2}(\phi)$ to the metric $g_{\mu \nu}$. Thus one deals with the Jordan frame metric $\tilde{g}_{\mu \nu}\equiv A^{2}(\phi) g_{\mu \nu}$ in the matter part of the action. Varying action (\ref{1}) with respect to the dynamical variables $g_{\mu \nu}$ and $\phi$ leads to the equations of motion:
\begin{equation}\label{2}
  8\pi M^{2}_{pl}G_{\mu \nu}=T^{\phi}_{\mu\nu} + A^{2}(\phi)\tilde{T}_{\mu \nu}
\end{equation}
and
\begin{equation}\label{3}
  \Box\phi=V_{,\phi}-A^{3}(\phi)A_{,\phi}(\phi)\tilde{T}=V_{,\phi}-\frac{\partial ln A(\phi)}{\partial \phi}T
\end{equation}
where $\tilde{T}_{\mu\nu}=\frac{-2}{\sqrt{-g}}\frac{\delta S_{M}}{\delta \tilde{g}_{\mu \nu}}$ and $\tilde{T}$ is its trace part. $T^{\phi}_{\mu \nu}$ has the usual definition:
\begin{equation}\label{4}
  T^{\phi}_{\mu \nu}=\partial_{\mu}\phi \partial_{\nu}\phi -\frac{1}{2}g_{\mu \nu} (\partial \phi)^{2}-g_{\mu \nu}V(\phi).
\end{equation}
The screening behaviour refers to the $\phi$-derivative of effective potential $V^{eff}_{,\phi}=V_{,\phi}-\frac{\partial ln A(\phi)}{\partial \phi}T$ in the Klein-Gordon equation (\ref{3}). In this model, the effects of the fifth force is suppressed against Newtonian gravitational force in dense environment while these effects reveal where the local density is rare. From the newtonian limit of the relations (\ref{2}) and (\ref{3}), one can conclude that
\begin{equation}\label{34}
F=-(G+\frac{d ln A(\phi)}{d \phi}\frac{\nabla \phi}{\nabla \Phi})\nabla \Phi
\end{equation}
is the effective gravitational force on a test particle of unit mass where $\Phi=\frac{\Phi_{N}}{G}$ is Newtonian potential per Newtonian constant. The first term is the Newtonian gravitational force and the second term indicates the effect of the fifth force on the particle. Therefore, one can roughly consider the effect of the fifth force in dense environment, at least in Newtonian limit, as a local and small modification of Newtonian constant $G_{N}$ \cite{zhang1,zhang2}. Through this paper we follow this idea and consider $G_{eff}=G_{N}+\delta G$ in thin shell of dense environment. By this approach we track the plausible effects of SMG in producing GW from a BH-BH binary system.

%----------------------------------------------
\section{BH-BH binary system in SMG}\label{BHBH}

Let's consider a binary system contains two astrophysical black holes, one massive black hole at the center of a spherically symmetric galactic ambient/halo with constant density $\rho_{0}$ \footnote{We follow the model with spherical galactic ambient/halo which has been used in \cite{ac1} as monopole approximation.}, and a small black hole as its companion. The black hole surrounded by either cold dark matter or scalar field dark mater halo was analytically studied by Xu et al. \cite{xu} in stationary situation. Their results show that dark matter halo make the black hole horizon to increase while one can recognize a small reduction in ergosphere. Further more using the Romulus25 simulation the authors in \cite{tremel} predicted that Milky Way-mass halos often host several (wandering)supermassive black holes which are likely to be found in galactic disk. So it may not be far from reality to consider that central black hole enfolded by galactic halo. Further central black hole is surrounded by an almost empty region which is known as ISCO region. ISCO is defined as a region that all the geodesics are towards the black hole. This means that all the particles in this region should fall into the black hole which result in an empty space. For Schwarzschild-like black holes one has $R_{ISCO}\simeq3R_{horizon}$. We focus on chameleon-like screening mechanism and rather than accretion disk a galactic ambient/halo around the black hole is considered. Therefore, one does not expect sharp change of scalar profile in vacuum region \cite{ac1,ac2} and the system shows mild screening. It is also assumed that the fifth force does not spoil the symmetry. Therefore, the density of background around the central black hole is given by the relation
\begin{equation}
            \begin{aligned}\label{6}
            \rho(r)=\begin{cases}0 & \text{$R_{H}<r <R_{ISCO}$} \\
            \rho_{0} & \text{$r> R_{ISCO}$} \\[1.2ex]
            \end{cases}
            \end{aligned}
            \end{equation}
where $R_{H}$ and $R_{ISCO}$ are the horizon and ISCO radii of central black hole, respectively. As the spatial-dependent densities around a central black hole is a promising area to look for the effects of the screening mechanism, we study the behaviour of the companion black hole in there. Indeed, with circular orbit approximation\footnote{This approximation is valid for $ r>R_{ISCO}$ \cite{mag}.}, we investigate the effect of fifth force on companion black hole when it crosses through thin shell in the vicinity of ISCO of central black hole. By thin shell we mean the region where the gradient of the scalar field makes the fifth force turns on. So the largest effect of the fifth force on dynamic of the binary system happens when the companion enters the thin shell of main black hole.

Consequently, we divide the region around central black hole to the three regions wherein gravitational constant takes the values
\begin{equation}\label{7}
            \begin{aligned}
            G_{eff}(r)=G_{N}\,\times\,\begin{cases}1 & \text{$R_{H}<r <R_{ISCO}$} \\
            1+\mu & \text{$R_{ISCO}<r<R_{thin shell}$} \\[1.2ex]
            1 & \text{$r>R_{thin shell}$}
            \end{cases}
            \end{aligned}
            \end{equation}
In this relation, $\mu\equiv\frac{\delta G}{G_{N}}$ which represents the leading order effects of SMG on the gravitational force.

\subsection{GW Emitted}

In this subsection, on the basis of the above mentioned assumptions, we drive the power of the gravitational wave of a binary system in the SMG framework. Before we going on, let us focus on conditions wherein no scalar mode emitted from the binary. In \cite{zhang1}, it is mentioned that the scalar mode in SMG could be excited only if the frequency of the scalar mode is greater than the scalar mass. Therefore, having no scalar mode translates to new constraints on the parameters of the model. As we want to focus only on the effects of SMG on the tensorial mode, we assume that in our model the frequency of scalar mode is less than scalar mass in the thin shell and so there is no excited scalar mode.

As mentioned earlier, we have considered the effect of fifth force as a modification of gravitational constant. Therefore, it is sufficient to substitute $G_{N}\rightarrow G(r)$ in the linearized Einstein equation of GR which leads to
\begin{equation}\label{20}
  \Box \bar{h}_{\mu \nu}=\frac{-16\pi}{c^4}  T^{eff}_{\mu \nu}
\end{equation}
where $\bar{h}_{\mu \nu}=h_{\mu \nu}-\frac{1}{2}\eta_{\mu \nu}h$. The tensor $h_{\mu \nu}$ is metric perturbation with the trace of $h$, and the effective energy-momentum tensor $T^{eff}_{\mu \nu}= G(r)T_{\mu \nu}$. In addition, as the effects of screening mechanism is small, we keep all the calculations up to the leading order of $\mu$. The solution of the relation (\ref{20}) up to the leading order of the metric perturbation $h_{\mu \nu}$ in traceless-transverse (TT) gauge is:
\begin{equation}\label{21}
  [h^{TT}_{\mu \nu}(t,\vec{x})]_{quadrupole}=\frac{2}{r c^4}\Lambda_{ij,kl}(\hat{n})\ddot{M}^{kl}_{eff}.
\end{equation}
The effective matter quadrupole for a binary system with masses $G_{1}m_{1}$ and $G_{2}m_{2}$ respectively at the position $\vec{x}_{1}$ and $\vec{x}_{2}$ can be written as
\begin{eqnarray}\label{22}
 {M}^{ij}_{eff} &&=m_{1}G_{1}x^{i}_{1}x^{j}_{1}+m_{2}G_{2}x^{i}_{2}x^{j}_{2} ,\nonumber\\
                &&= (m_{1}G_{1}+m_{2}G_{2})x^{i}_{CM,eff}x^{j}_{CM,eff}+\mathcal{M}_{eff}(x^{i}_{CM,eff}x^{j}_{0}+x^{j}_{CM,eff}x^{i}_{0})+\mathcal{M}_{eff}x^{i}_{0}x^{j}_{0}
\end{eqnarray}
where the effective center of mass and effective reduced mass and separation vector are defined respectively as:
\begin{eqnarray}\label{23}
  && \vec{X}^{eff}_{CM}\equiv\frac{G_{N}\,m_{1}\,\vec{x}_{1}+(G_{N}+\delta G)\,m_{2}\,\vec{x}_{2}}{G_{N}\,m_{1}+(G_{N}+\delta G)\,m_{2}} ,\nonumber\\
  &&\mathcal{M}^{eff} \equiv \frac{G_{N}\,m_{1}\,(G_{N}+\delta G)\,m_{2}}{G_{N}\,m_{1}+(G_{N}+\delta G)\,m_{2}},\nonumber\\
  &&\vec{X}_{0}\equiv\vec{X}_{1}-\vec{X}_{2} .
\end{eqnarray}
Note that all of the above relations reduce to the standard ones in the absence of the modified gravity i.e. $\delta G\rightarrow 0$. In term of leading order of $\mu$, the first two relations in (\ref{23}) take the form of:
\begin{eqnarray}\label{24}
  && \vec{X}^{eff}_{CM}\simeq \vec{X}_{CM}-\mu \frac{\mathcal{M}}{m_{1}+m_{2}}\vec{X}_{0}+\mathcal{O}(\mu^{2}) ,\nonumber \\
  &&\mathcal{M}^{eff} \equiv G_{N}\mathcal{M}(1+\mu\frac{m_{1}}{m_{1}+m_{2}})+\mathcal{O}(\mu^{2})
\end{eqnarray}
where $\mathcal{M}\equiv\frac{m_{1}m_{2}}{m_{1}+m_{2}}$ is the reduced mass. By substituting relations (\ref{23}) and (\ref{24}) in to the relation (\ref{22}) and keeping terms to the first order of $\mu$, one has
\begin{eqnarray}\label{25}
{M}^{ij}_{eff} & = & G_{N}\bigg\{(m_{1}+m_{2})x^{i}_{CM}x^{j}_{CM}+\mathcal{M}(x^{i}_{CM}x^{j}_{0}+x^{j}_{CM}x^{i}_{0}+x^{i}_{0}x^{j}_{0})\nonumber \\
&&+\mu\bigg[m_{2}x^{i}_{CM}x^{j}_{CM}-\mathcal{M}(m_{1}+m_{2})(x^{i}_{CM}x^{j}_{0}+x^{j}_{CM}x^{i}_{0})
 + \mathcal{M}\frac{m_{1}}{m_{1}+m_{2}}(x^{i}_{CM}x^{j}_{0}+x^{j}_{CM}x^{i}_{0}+x^{i}_{0}x^{j}_{0}) \nonumber \\
   &&+(\frac{\mathcal{M}^2}{m_{1}+m_{2}})x^{i}_{0}x^{j}_{0}-2\mathcal{M} \bigg] \bigg\}.
\end{eqnarray}
Choosing center of mass frame of reference, (\ref{25}) reduces to the simple relation
\begin{equation}\label{26}
 {M}^{ij}_{eff}=G_{N}\mathcal{M}x^{i}_{0}x^{j}_{0}(1+\mu \lambda_{m})
\end{equation}
where we have introduced $\lambda_{m}\equiv\frac{1-m_2/m_1}{(1+m_{2}/m_1)^{2}}$. Therefore, it is evident that to have a non-vanishing $\lambda_{m}$ and observing the effects of SMG, the two objects of binary system must have different masses. The standard GR counterpart of the relation (\ref{26}) is
\begin{equation}\label{26a}
 {M}^{ij}=G_{N}\mathcal{M}x^{i}_{0}x^{j}_{0}
\end{equation}
 which means that  SMG only modifies  constant coefficient $G_{N} \mathcal{M}$ to $G_{N}\mathcal{M}(1+\mu \lambda_{m})$. In addition, the power carried by GW in GR is given by the relation
\begin{equation}\label{18}
  P=\frac{32 c^5}{5}\frac{1}{G_{N}}(\frac{G_{N}\,M_{c}\,\omega_{GW}}{2c^3})^{10/3}
\end{equation}
where $M_{c}\equiv \mathcal{M}^{\frac{3}{5}}m^{\frac{2}{5}}$ is the chirp mass and $ m= m_{1}+m_{2}$ is total mass. Consequently, it is easy to show that the carried power of GW in SMG, to the leading order of $\mu$, will be:
\begin{equation}\label{27}
  \tilde{P}=\frac{32 c^5}{5}\frac{1}{G_{N}}(\frac{G_{N}\,\tilde{M_{c}}\,\omega_{GW}}{2c^3})^{10/3}
\end{equation}
where $\tilde{M_{c}}\equiv M_{c}(1+\mu \lambda_{m})^{7/10}$.
%----------------------------------------------
\subsection{Energy loss of the spiral }

In the previous subsection, we have derived the power carried by the gravitational emission. This power should be the same as the energy loss of the binary system. Before evaluating the energy loss of the orbit of the binary system in SMG, let's review the concept of quasi-circular orbit approximation in GR. The source of the radiated energy for two point-like objects $m_{1}$ and $m_{2}$ with orbital radius R is defined as total energy of the orbit, namely \cite{mag}
\begin{equation}\label{8}
E_{orbit}=-\frac{G_{N}m_{1}m_{2}}{2R}.
\end{equation}
The binary is described in Newtonian limit and third Kepler's law is valid. Considering $\omega_{s}$ as the orbital frequency and $m$ as the total mass of the system, one has
\begin{equation}\label{9}
 \omega^{2}_{s}=\frac{G_{N}m}{R^{3}}.
\end{equation}
Time derivative of both sides of relation (\ref{9}) shows that if $\dot{\omega}_{s}\ll\omega^{2}$, then $\dot R\ll \omega_s R$. This condition is called quasi-circular orbit approximation.

To rewrite the above relations for our model in SMG, let's consider $G_{N}\longrightarrow G(r)$ according to the relation (\ref{7}). Hence, the total orbital energy will be
\begin{equation}\label{13}
  \tilde{E}_{orbit}=-G(r)\frac{m_{1}m_{2}}{2R}.
\end{equation}
Time derivative of both sides of this relation results
\begin{equation}\label{14}
  \frac{d\tilde{E}_{orbit}}{dt}=-G_{N}\frac{m_{1}m_{2}}{2R^{2}}\dot{R}\big(1-\eta R\big).
\end{equation}
which is the energy loss of the binary system during circular motion with a non-constant $G(r)$. In this relation we have introduced a parameter $\eta\equiv\frac{G'}{G}$ where $G'=\frac{dG(r)}{dr}$. By assuming that the variation of gravitational coupling constant in thin shell region is slow, which is in agreement with our previous assumption for the system to be in the galactic halo/ambient \cite{ac1}, one gets $\eta\equiv\frac{G'}{G}\ll 1$. It is evident that this parameter is related to the $\mu$ according to $\eta = \frac{\mu}{\Delta R}$ where $\Delta R$ indicates the thickness of the thin shell. Note that we will consider $\eta R \ll 1$ which corresponds to $\mu \ll \frac{\Delta R}{R}$; this could guarantee mild screening effect where we expect in thin shell for black hole in the galactic halo/ambient \cite{ac1}.

Kepler's third law in our model is written as
\begin{eqnarray}\label{14aa}
\tilde{\omega}^{2}_{s}&=&\frac{G_{N}m_{1}+(G_{N}+\delta G)m_{2}}{R^{3}}\nonumber\\
&=&\frac{G_{N}m}{R^{3}}(1+\mu \frac{m_{2}}{m})
\end{eqnarray}
which leads to
\begin{equation}\label{15}
  R\simeq (\frac{G_{N}m}{\tilde{\omega}^{2}_{s}})^{\frac{1}{3}}(1+\mu \frac{m_{2}}{3m}).
  \end{equation}
  From this one has the relations
 \begin{eqnarray}\label{16}
  &&\dot{R}\simeq -\frac{2}{3}\dot{\tilde{\omega}}_{s}(\frac{G_{N}m}{\tilde{\omega}^{5}_{s}})^{\frac{1}{3}}(1+\mu \frac{m_{2}}{3m} )\\
  &&R^{-2}\simeq (\frac{G_{N}m}{\tilde{\omega}^{2}_{s}})^{-\frac{2}{3}}(1-\mu \frac{2m_{2}}{3m})\label{16a},
\end{eqnarray}
to the leading order of $\mu$. Inserting relations (\ref{15}-\ref{16a}) in (\ref{14}) and using $\omega_{GW}=2\tilde{\omega}_{s}$ leads to
\begin{equation}\label{17}
  \frac{d\tilde{E}_{orbit}}{dt}\simeq -\frac{2}{3}(\frac{G_{N}^{2}M_{C}^{5}}{32})^{1/3}\dot{\omega}_{GW}\omega_{GW}^{-1/3}(1-\mu \frac{m_{2}}{3m}).
\end{equation}
which is the energy loss of the orbit in terms of GW orbital frequency and up to the leading order of SMG modifications.\footnote{Note that in \cite{ac1} there is also an estimation on the effects of fifth force in the power carried by the radiation, although without any explicit calculations.} We have used condition $\eta R\ll 1$ and circular orbit approximation $\dot{\omega}_{s}\ll \omega^{2}_{s}$ to eliminate $\eta$-term. In the next subsection we will find the relation between time and frequency of the GW in the presence of SMG which can be used as a clue to look for screening effects in GW data.

\subsection{GW frequency}
By combining (\ref{17}) and (\ref{27}), one can derive the differential equation for orbital GW frequency which is:
\begin{equation}\label{28}
  \dot{\omega}_{GW}=\frac{12}{5}(\frac{32 G_{N}M_{c}}{c^3})^{5/3}(1+\mu \tilde{\lambda}_{m})\omega_{GW}^{11/3}
\end{equation}
up to the leading order of $\mu$. In this relation $\tilde{\lambda}_{m}=\frac{7}{3}\lambda_{m}+\frac{m_{2}}{3m}$. As the wave frequency is defined as  $f_{GW}=\frac{\omega_{GW}}{2\pi}$, one has
\begin{equation}\label{29}
  \dot{f}_{GW}=\frac{96}{5}\pi^{8/3}(\frac{G_{N}M_{c}}{c^3})^{5/3}(1+\mu \tilde{\lambda}_{m})f^{11/3}.
\end{equation}
Integrating (\ref{29}) shows that $ \dot{f}_{GW}$ diverges at a finite coalescence time, denoted by $t_{coal}$. Therefore, by introducing $\tau=t_{coal}-t$, we get
\begin{eqnarray}\label{30}
  f_{GW}(\tau)&\simeq&\frac{1}{\pi}(\frac{G_{N}M_{c}}{c^3})^{-5/8}(\frac{5}{256\tau})^{3/8}(1-\frac{3}{8}\tilde{\lambda}_{m}\mu)\nonumber\\
  &=& 134 Hz (\frac{1.21 M_{sun}}{M_c})^{5/8} (\frac{1 s}{\tau})^{3/8} (1-\frac{3}{8}\tilde{\lambda}_{m}\mu)
\end{eqnarray}
to the leading order of $\mu$. This relation gives information about the frequency of the emitted wave once the companion black hole enters the thin shell region, where the effects of fifth force turn on. Note that for the radius greater that $R_{thinshell}$ and smaller than $R_{ISCO}$, the results of standard GR are valid. From (\ref{30}), it is clear that, because of the parameter $\mu$, the frequency of the gravitational emission have decreased in comparison with GR. Let's define relative change in GW frequency between the two cases, with and without screening
 \begin{equation}
f_{R}=\frac{f_{GW}[\mu=0]-f_{GW}[\mu]}{f_{GW}[\mu=0]}=\frac{3}{8}\tilde{\lambda}_{m}\mu
\end{equation}
which is a linear function of $\mu$ and is independent of $\tau$. If we consider a binary with condition $\frac{m_2}{m_1} \rightarrow 0$, relative change in GW frequency $f_{R}$ will be $\frac{7}{8} \mu$. %Figure (\ref{j}) illustrates the plot of $f_{R}$ for a binary comprises of two masses of $m_{1}$ and $m_{2}$ and we have considered a fixed small fifth force effect and set it as $\mu = 10^{-4}$.

%\begin{figure}
%	\centering\epsfig{file=J1.pdf,width=8cm,angle=0}\epsfig{file=J2,width=8cm,angle=0}
%	\caption{\footnotesize The left plot shows relative change in GW frequency for $\mu=10^{-4}$ while the right one present corresponding $\tau_{R}$.}
%\end{figure}

\begin{figure}\label{j}
\centering\epsfig{file=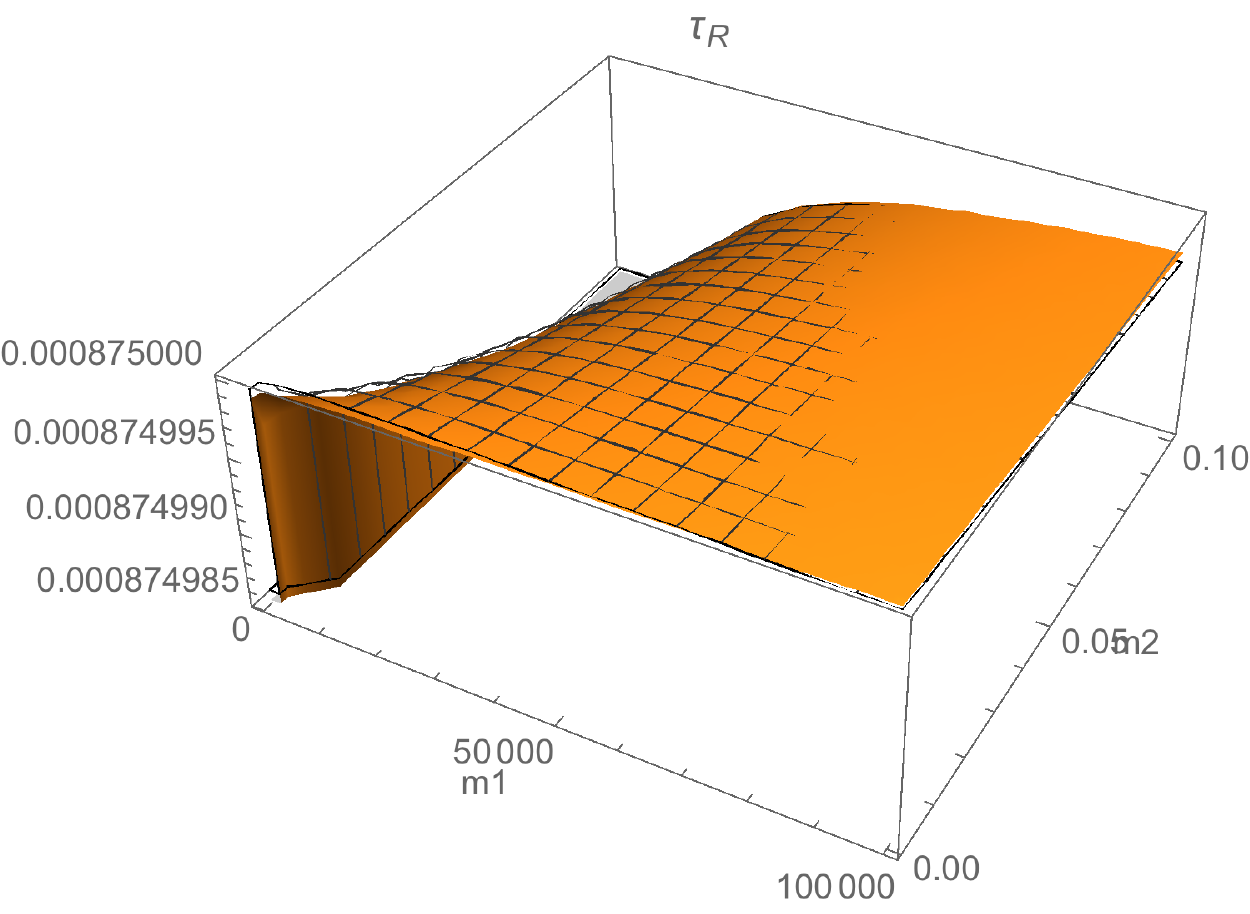,width=9cm,angle=0}
\caption{\footnotesize The plot illustrates $\tau_{R}$ in terms of $m_1$ and $m_2$ for $\mu=10^{-4}$. }\label{j2}
\end{figure}

In addition, relation (\ref{30}) shows that because of the parameter $\mu$ the time of coalescence is increased and the fifth force can be seen as an extra attraction between the companion black hole and the environment. From (\ref{14aa}) we can derive a relation for the frequency of the wave which has been emitted when the companion black hole reached the ISCO radius. We have considered a black hole with mass $m$ at the center, so $R_{ISCO}\simeq 3R_{H}=\frac{6G_{N}m}{c^2}$. Therefore:

\begin{equation}\label{31}
 f^{ISCO}_{GW}=\frac{c^3}{6\pi\sqrt{6}}\frac{1}{mG_{N}}\big(1+\mu\frac{m_{2}}{2m}\big)= 2.2(kHz)\, \frac{M_{sun}}{m}\big(1+\mu\frac{m_{2}}{2m}\big).
\end{equation}
Substituting above frequency in to the relation (\ref{30}) we obtain the $\tau$ related to the $R_{ISCO}$. We can also compute relative change in $\tau$ which is
 \begin{equation}\label{32}
\tau_{R}=\frac{\tau [\mu=0]-\tau[\mu]}{\tau[\mu=0]}=\frac{3}{8}\big(\tilde{\lambda}_{m}+\frac{m_2}{m}\big)\mu.
\end{equation}
Figure (\ref{j2}) shows an illustration of $\tau_R$  in terms of $m_1$ and $m_2$ for $\mu=10^{-4}$ when the companion black hole reaches to $r=R_{ISCO}$.

%From the figure, one can see when $\frac{m_{1}}{m_{2}}$ is a bout $10^8$, the fifth force cause a longer time for coalescence a bout 2.5 s.

It is instructive to find the relation between $\mu=\frac{\delta G}{G_N}$ which has been used above and $A$, the original parameter of chameleon model in (\ref{34}). For this purpose, we assume the scalar potential $V(\phi)= M^{4+n}\phi^{-n}$ and the coupling  $A(\phi)= e^{\beta \phi/ M_{pl}}$ where $n$ and $\beta$ are integers of order one. The mass scale $M$ is constrained by local experiments $M \lesssim 10^{-3}eV$ \cite{smcos,ac1}. By using relation (\ref{34}) it is easy to show that
\begin{eqnarray}\label{33}
\mu &=& |\frac{d \ln A}{d\phi}.\frac{\nabla \phi}{G\nabla \Phi}|_{R_{ISCO}}\nonumber\\
&\approx& (\frac{R_{ISCO}}{R_{H}})^{2}\beta |\nabla \phi|_{R_{ISCO}}\frac{m_{1}}{M_{pl}^{3}}.
\end{eqnarray}
In \cite{ac1}, it is shown that for the chameleon model the scalar gradient near ISCO can be written as
\begin{equation}\label{35}
|\nabla \phi|_{R_{ISCO}}\approx \frac{\beta \rho_{0}R_{ISCO} }{2M_{pl}}
\end{equation}
where $\rho_{0}$ is the density of a galactic halo. By substituting (\ref{35}) into (\ref{33}) we get
\begin{eqnarray}\label{36}
\mu \approx {\beta^2}\rho_{0} \frac{\,m_{1} R_{ISCO}^3}{2M_{pl}^{4} R_{H}^{2}}
    \approx 27 \beta^{2}\rho_{0}\frac{\,m_{1}^{2}}{M_{pl}^{6}}
\end{eqnarray}
which shows that the relation between $\mu$ and $\beta$ depends on the environment. So if we know the host of binary then GW can be used to put constraints on the $\beta$ parameter of the chameleon model. However, the above relation shows that $\mu$ does not depend on the mass scale $M$.

\section{Summary and Conclusions}
The physics of gravitational waves becoming very important in the community due to the recent direct observation of GW of binaries \cite{ligo}. In this work we have studied the effects of modification of gravity in the GW pattern. Note that it has been shown in the literature that the existence of any modified gravity is well constrained. This is claimed by comparing the propagation speed of GW and its electromagnetic counterpart from  NS-NS GW observation. But we should emphasize that here we focus on the modified gravity models that present the screening mechanism. We have shown that the screening mechanism affect the frequency of GW just before the collapse of the binaries. The reason for this result is modification of gravity law in the ISCO region.

To observe the above effect on GW pattern we need very clean data just before the collapse of binaries. This needs more and more GW event observations due to very smallness of the screening mechanism if it exists. Fortunately, LISA and LIGO/Virgo will run for observing more and more GW from binaries in different mass regimes. In this work we have done our analysis analytically by employing approximations but in the future we need to do accurate simulations of BH-BH binaries in the presence of screening mechanism.

\section*{Acknowledgments}
We would like to thank A.C. Davis for her comments on the draft.

\end{document}